\documentclass[letterpaper,twocolumn,english,prl,superscriptaddress]{revtex4-1}
\usepackage[T1]{fontenc}
\usepackage[latin9]{inputenc}
\usepackage{color}
\usepackage{amsmath}
\usepackage{amssymb}
\usepackage{graphicx}
\usepackage{esint}


\usepackage{graphicx}

\makeatother

\usepackage{babel}
\begin{document}

\title{Chiral edge plasmons in quantum anomalous Hall insulators}

\author{Furu Zhang}

\thanks{These authors contributed equally to this work.}

\affiliation{School of Science, Beijing Forestry University, Beijing 100083, China}

\affiliation{Key Laboratory of Advanced Optoelectronic Quantum Architecture and
Measurement (MOE), School of Physics, Beijing Institute of Technology,
Beijing 100081, China}

\author{Chenxi Ding}

\thanks{These authors contributed equally to this work.}

\affiliation{\textcolor{black}{Anhui Key Laboratory of Low-Energy Quantum Materials
and Devices, High Magnetic Field Laboratory, HFIPS, Anhui, Chinese
Academy of Sciences, Hefei 230031, P. R. China}}

\affiliation{University of Science and Technology of China, Hefei, 230026, P. R. China}

\author{Jianhui Zhou}
\email{jhzhou@hmfl.ac.cn}

\affiliation{\textcolor{black}{Anhui Key Laboratory of Low-Energy Quantum Materials
and Devices, High Magnetic Field Laboratory, HFIPS, Anhui, Chinese
Academy of Sciences, Hefei 230031, P. R. China}}

\author{Yugui Yao}
\email{ygyao@bit.edu.cn}

\affiliation{Key Laboratory of Advanced Optoelectronic Quantum Architecture and
Measurement (MOE), School of Physics, Beijing Institute of Technology,
Beijing 100081, China}

\begin{abstract}
We find that the Berry curvature splits the edge plasmons propagating
along the opposite directions in quantum anomalous Hall insulators
even with vanishing Chern number. When the bulk is insulating, only
one unidirectional edge plasmon mode survives whose direction can
be changed by external fields. The unidirectional edge plasmon in
the long-wavelength limit is acoustic and essentially determined by
the anomalous Hall conductivity. The group velocity of the chiral
edge plasmon would change its sign for a large wave vector, which
originates from the $k$-quadratic correction to the effective mass.
The impacts of the Fermi level and the wave vector on the bulk and
edge plasmons are discussed. Our work provides a well quantitative
explanation of the recent observation of the chiral edge plasmon in
quantum anomalous Hall insulators and some insight into the application
of realistic topological materials in chiral plasmonics. 
\end{abstract}
\maketitle
\textit{Introduction.-{}-}Berry phase is a key ingredient of modern
quantum theory and plays an important role in the development of topological
phases of matter and topological materials \citep{Xiao10RMP}. Recently,
the Berry phase effect on the plasmon excitations in solids has attracted
increasing attentions \citep{Juergens2014PRL,Zhou2015Plasmon,Kumar2016PRB,Song2016pnas,ZhangFR2017PRL,Zhang2018PRB,Cao2021PRL,Islam2021PRB,Heidari2021PRB,LiangZ2023PRB}.
Several theoretical works have shown that the Berry curvature, an
effective magnetic field in momentum space, should split the energy
spectra of edge plasmons in the metallic systems such as photo-excited
two-dimensional (2D) gapped Dirac materials at zero external magnetic
fields \citep{Kumar2016PRB,Song2016pnas}. Unlike the conventional
magnetoplasmons \citep{Fetter1985PRB,Fetter1986PRB1,Volkov1988jetp,Bartolomei2023PRL,Sokolik2024PRB},
magnetic fields are not need to break the time reversal symmetry and
to rise a definite chirality of magnetoplasmons. In addition, the
chiral plasmons induced by the Berry curvature are themselves handed
and essentially differ from those ones in either metallic handed nanoparticles
or handed groupings of individual resonant nanoparticles for chiral
plasmonics \citep{Hentschele2017SA}. 

Topological materials including topological insulators (TIs) \citep{Hasan2010RMP,QiZhangRMP}
and topological semimetals \citep{Armitage2018RMP} provide us promising
platforms to investigate the impacts of the topology of the Bloch
bands on collective excitations \citep{maier2010plasmonics,grigorenko2012NPho}.
Recently there have been a series of important experimental progress
on plasmon excitations in the topological materials \citep{diPietro2013NN,ou2014NC,Kogar2015PRL,Glinka2016NC,JiaX2017PRL,Politano2018PRL,Xue2021PRL,WangC2021PRA,Shao2022SciAdv,LiY2023PRB}.
In particular, the contactless microwave circulator response suggests
the existence of the chiral edge plasmons in magnetized disks of TIs,
Cr-doped $\left(\mathrm{Bi},\mathrm{Sb}\right)_{2}\mathrm{Te}_{3}$,
that support the quantum anomalous Hall effect (QAHE) \citep{Mahoney2017NC,WangTG2023PRB,Martinez2024PRR}.
These chiral edge plasmons exhibit a strong magnetic-free nonreciprocity
that enables us to design and implement nonreciprocal devices in photonics
\citep{Feng729Science} and transceiver technology \citep{Bi2011NP,Estep2014NP,Mahoney2017PRX}.
Unlike the conventional band insulators, the QAH insulators host robust
chiral edge states \citep{Haldane1988PRL,Yu2010Science,changCZ2013Science,Bestwick2015QAHE,WengQAHE2015,ChangCZ2023RMP},
which may support the chiral edge plasmons. However, a thorough and
direct theoretical explanation of the chiral edge plasmons in QAH
insulators as well as their manipulation remain unexplored. 

In this work, we show that the Berry curvature splits the energy dispersion
of edge plasmons in the magnetically doped thin films of TIs. When
the bulk is insulating, only one unidirectional edge plasmon mode
survives, whose direction can be changed by external fields. The unidirectional
edge plasmon in the long-wavelength limit is acoustic and entirely
determined by the quantum anomalous Hall conductivity and the effective
dielectric constant of the environment. For a large wave vector, the
group velocity of the chiral edge plasmon would change its sign, which
originates from the $k$-quadratic correction of the effective mass.
The behaviors of both the bulk and edge plasmons dependent on the
Fermi level and the wave vector are discussed. 

\begin{figure}[b]
\begin{centering}
\includegraphics[scale=0.9]{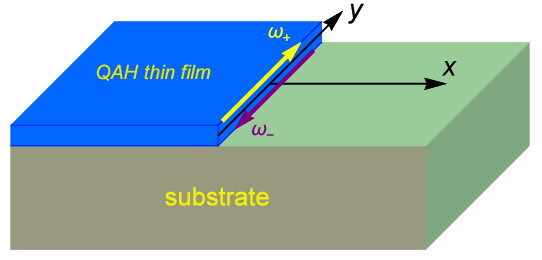}
\par\end{centering}
\centering{}\caption{A schematic drawing of the physical model of edge plasmons $\omega_{\pm}$
near the edge between the QAH insulator and a vacuum. The 2D semi-infinite
QAH system is put on the substrate, occupying the region $x<0$. \label{fig1} }
\end{figure}

\textit{Formalism of edge plasmons}.\textit{-{}-}We begin with a physical
model of the system, in which a 2D semi-infinite QAH system occupying
the region $x<0$ is put on the substrate, as shown in Fig. \ref{fig1}.
The effective dielectric constant of the substrate ($z<0$) is $\epsilon_{sub}$,
and there is a vacuum for $z>0$. We consider an edge along the $y$
direction. The edge plasmon excitations are governed by a set of self-consistent
equations \citep{WuJW1986PRB}: the Poisson's equation $\nabla\cdotp[\varepsilon E(\boldsymbol{r})]=\rho(x,y)\delta(z),$
and the equation of continuity $\nabla\cdot j(x,y)=i\omega\rho(x,y)$,
where $\rho(x,y)=\rho(x)e^{i(qy-\omega t)}$ is the charge density
fluctuation and $j_{\alpha}(x,y)=\underset{\beta}{\sum}\sigma_{\alpha\beta}E_{\beta}(x,y,z=0)$
is the current density. The frequency and wave vector dependent electrical
conductivity tensor $\sigma_{\alpha\beta}(q,\omega)$ is calculated
through the general Kubo formula \citep{giuliani2005qtel}. $\varepsilon=\epsilon_{0}$
for $z>0$ and $\varepsilon=\epsilon_{sub}\epsilon_{0}$ for $z<0$.
In this work, we use the average surrounding dielectric constant $\epsilon=(1+\epsilon_{sub})\epsilon_{0}/2=7\epsilon_{0}$
to directly compare with the recent experiment, in which the QAH insulator
is grown on the substrate of semi-insulating GaAs \citep{Mahoney2017NC}. 

The dispersion relations of plasmons can be obtained from solving
the set of self-consistent equations above. By means of the Laguerre
series \citep{WuJW1986PRB,WangWH2011PRB}, one gets the numerical solutions in a matrix equation: 
\begin{equation}
\det\left[J_{mn}+(\eta_{1}\pm\chi)I_{m}-\delta_{mn}\right]=0,\label{DetEq}
\end{equation}
where the parameters are $\eta_{1}=\frac{|q|\sigma_{xx}}{i\epsilon\omega}$,
$\eta_{2}=\frac{|q|\sigma_{yy}}{i\epsilon\omega}$ and $\chi=\frac{|q|\sigma_{xy}}{\epsilon\omega}.$
We leave the specific expressions of $J_{mn}$ and $I_{m}$ given
in the Supplemental Material \citep{SMCEP}. 

After choosing the approximate expression of the integral kernel \citep{Fetter1985PRB,Fetter1986PRB1},
one can analytically solve the self-consistent equations and yield
the corresponding dispersion relation of the edge plasmons \citep{SMCEP} 
\begin{equation}
\eta_{1}\eta_{2}-2\eta_{1}-\eta_{2}-\chi^{2}\pm2\sqrt{2}\chi=0,\label{AppEq}
\end{equation}
where the plus (minus) sign denotes that the edge plasmon propagates
along the $y\left(-y\right)$ direction. From Eq. $\left(\ref{AppEq}\right)$,
one clearly sees that a non-zero transverse conductivity or anomalous
Hall conductivity $\sigma_{xy}$ breaks the degeneracy of the edge
plasmons propagating in opposite directions, yielding a pair of chiral
plasmons. For the time reversal invariant system, the anomalous Hall
conductivity $\sigma_{xy}$ should vanish. Thus the two edge plasmon
modes are degenerate. Note that taking $\eta_{1}=\eta_{2}$, Eq. $\left(\ref{AppEq}\right)$
would reduce to  the counterpart for the edge magnetoplasmons in 2D
electron gases \citep{Mast1985PRL,WuJW1985PRL}. 

\textit{Chiral plasmons in doped QAH insulators}.\textit{-{}-}Let
us consider the behaviors of plasmons near the edge of the QAH insulators.
To demonstrate the main physics, we choose the effective model for
QAH insulators in magnetically doped thin films of three-dimensional
strong TIs V- and Cr-doped $\left(\mathrm{Bi},\mathrm{Sb}\right)_{2}\mathrm{Te}_{3}$
\citep{LuHZ2013PRL,ZhangFR2017PRL}: 
\begin{equation}
H=H_{0}+\frac{m}{2}\tau_{0}\otimes\sigma_{z},\label{Ham}
\end{equation}
where $m$ is the exchange field originating from the magnetic dopants,
effectively acting as a Zeeman field. $H_{0}$ is given as \citep{zhang2009TI}
\begin{equation}
H_{0}=-Dk^{2}+\left(\begin{array}{cc}
h_{+}\left(\boldsymbol{k}\right) & V\\
V & h_{-}\left(\boldsymbol{k}\right)
\end{array}\right),\label{H0}
\end{equation}
with 
\begin{equation}
h_{\pm}\left(\boldsymbol{k}\right)=\left(\begin{array}{cc}
\pm(\frac{\Delta}{2}-Bk^{2}) & iv_{F}k_{-}\\
-iv_{F}k_{+} & \mp(\frac{\Delta}{2}-Bk^{2})
\end{array}\right),\label{Hpm}
\end{equation}
where $h_{\pm}\left(\boldsymbol{k}\right)$ describes the 2D Dirac
fermions with a $k$-dependent mass. $\boldsymbol{k}=(k_{x},k_{y})$
is the 2D wave vector and $k_{\pm}=k_{x}\pm ik_{y}$. $v_{F}$ is
the effective velocity. The $D$ term breaks the particle-hole symmetry
and splits the longitudinal bulk plasmon modes \citep{ZhangFR2017PRL}.
$\Delta$ is the hybridization of the top and bottom surface states
of the thin film. $V$ measures the structural inversion asymmetry
between the top and bottom surfaces. Pauli matrices $\tau_{0}$ and
$\sigma_{z}$ act on the pseudospin space related to the top and bottom
surfaces and the real spin degree of freedom, respectively. Here we
primarily consider the insulating phase that requires $\left|D\right|<\left|B\right|$.
In addition, we take the parameters in real calculations such that
the theoretical calculations of energy bands are in a good agreement
with experimental observations \citep{Yu2010Science,changCZ2013Science}.
Specifically, $v_{F}=3.0\:\mathrm{eV}\cdot\mathbf{\mathrm{\mathring{A}}}$,
$\Delta=-0.01\:\mathrm{eV}$ and $B=-30\:\mathrm{eV}\cdot\mathbf{\mathrm{\mathring{A}}}^{2}$. 

\begin{figure}
\begin{centering}
\includegraphics[scale=0.8]{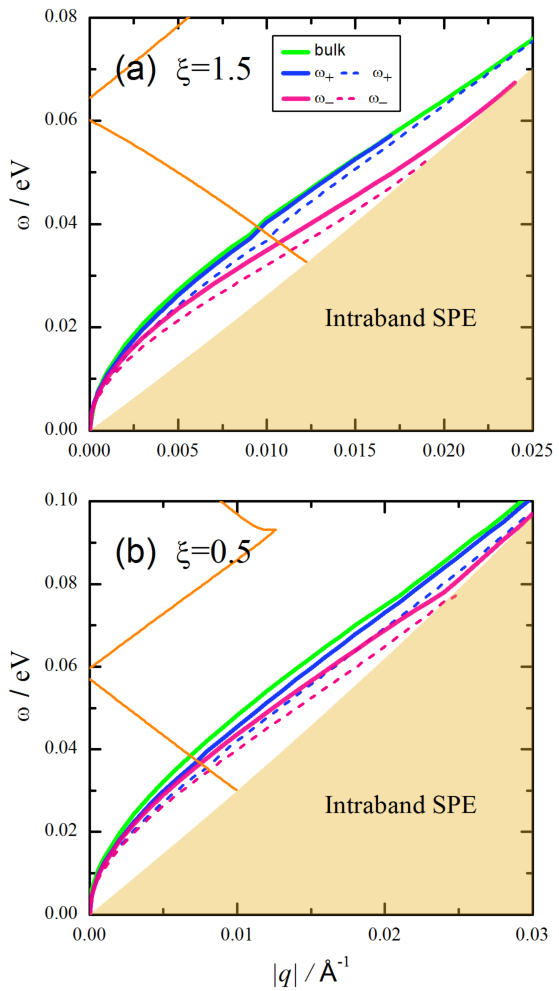}
\par\end{centering}
\centering{}\caption{The bulk plasmon (green lines) and chiral edge plasmons (blue and
purple lines) of semi-infinite QAH insulators in different topological
phases: $\xi=1.5$ for a nontrivial phase with a unit Chern number
in (a) and $\xi=0.5$ for a trivial phase with a vanishing Chern number
in (b). The orange lines refer to the boundary of the interband SPEs.
The solid (dashed) line indicates the numerical exact (approximate)
solutions. The parameters are $\mu=0.09\:\mathrm{eV}$ and $V=0.03\:\mathrm{eV}$,
$D=10\:\mathrm{eV}\mathrm{\mathring{A}^{2}}$. \label{fig2} }
\end{figure}

By the numerical solutions to Eq. $\left(\ref{DetEq}\right)$ and
the approximate analytical results from Eq. $\left(\ref{AppEq}\right)$,
we obtain the plasmon dispersions in doped QAH insulators in both
the topologically trivial ($\xi\equiv-m/m_{0}<1$ with $m_{0}=\sqrt{4V^{2}+\Delta^{2}}$)
and nontrivial $\left(\xi>1\right)$ phases, as plotted in Fig. \ref{fig2}.
The plasmon modes possess several key features. First, there are two
edge plasmon modes propagating in opposite directions near the edge
for both the topological nontrivial (Fig. \ref{fig2}(a)) and trivial
(Fig. \ref{fig2}(b)) phases. It is a sharp contrast to the previous
theoretical work on metallic systems that requires a nonzero anomalous
Hall conductivity or mean Berry curvature to split the edge plasmon
into a pair of chiral edge plasmons \citep{Kumar2016PRB,Song2016pnas,ZhangYa2017PRB}.
We define the mode propagating along the $y$ direction as $\omega_{+}$,
and the opposite mode as $\omega_{-}$. Second, Fig. \ref{fig2} shows
that the two edge plasmons are not degenerate and the energy of mode
$\omega_{+}$ is higher than mode $\omega_{-}$. Third, the numerical
results are well consistent with the corresponding approximate analytical
solutions, and the energy of the former is slightly higher \citep{WangWH2012PRB}.
Fourth, in contrast to the conventional magnetoplasmons in the context
of 2D electron gases \citep{Fetter1985PRB,WuJW1986PRB,Fetter1986PRB1},
both of the two edge plasmons are gapless. When the wave vector $q$
is small, $\omega_{+}$ and $\omega_{-}$ are nearly degenerate. As
the wave vector $q$ increases, the split of the two edge plasmon
modes becomes pronounced. The energy of the edge plasmons is lower
than the bulk plasmon that is determined by the zeros of the complex
dielectric function $\varepsilon(q,\omega)$ \citep{ZhangFR2017PRL},
$i.e.$ $\omega_{bulk}>\omega_{+}>\omega_{-}$. In the direction perpendicular
to the edge of the material, mode $\omega_{-}$ decays faster than
$\omega_{+}$. For a sufficiently large wave vector $q$, the fast
mode $\omega_{+}$ will merge with bulk mode $\omega_{bulk}$. Consequently,
only the edge plasmon $\omega_{-}$ survives along the edge until
it enters into the intraband single-particle excitations (SPEs) region.
In fact, the interband SPEs have little influence on the plasmons
because of tiny overlap of energy bands for the frequency range we
discussed here. In addition, comparing Figs. \ref{fig2}(a) with \ref{fig2}(b),
the behavior of the chiral plasmons in different phases are similar,
while the energy splitting of $\omega_{\pm}$ in the nontrivial phase
is much bigger than those in the trivial phase. 

\textit{Unidirectional plasmons in intrinsic systems}.\textit{-{}-}When
the bulk is insulating, the bulk plasmon $\omega_{bulk}$ will disappear
due to lack of free bulk electrons at the Fermi level. But the edge
plasmon still exists and is unidirectional. From the perspective of
semiclassical dynamics, the Berry curvature $\boldsymbol{\Omega}(\boldsymbol{k})$
acts as an effective magnetic field in momentum space and modifies
the kinematic equation of electrons with an anomalous velocity $-\text{\ensuremath{\frac{e}{\hbar}\boldsymbol{E}\times\boldsymbol{\Omega}(\boldsymbol{k})}}$,
which essentially accounts for various anomalous transport and optical
properties of Bloch electrons in solids \citep{Xiao10RMP}. The sign
of the Berry curvature determines the direction of the anomalous velocity
then significantly affects the behaviors of edge plasmons. Specifically,
for a nonzero wave vector $q$, there exists a single unidirectional
edge plasmon propagating on the edge of the magnetically doped TIs,
even when it is topologically trivial. 

\begin{figure}[b]
\begin{centering}
\includegraphics[scale=0.7]{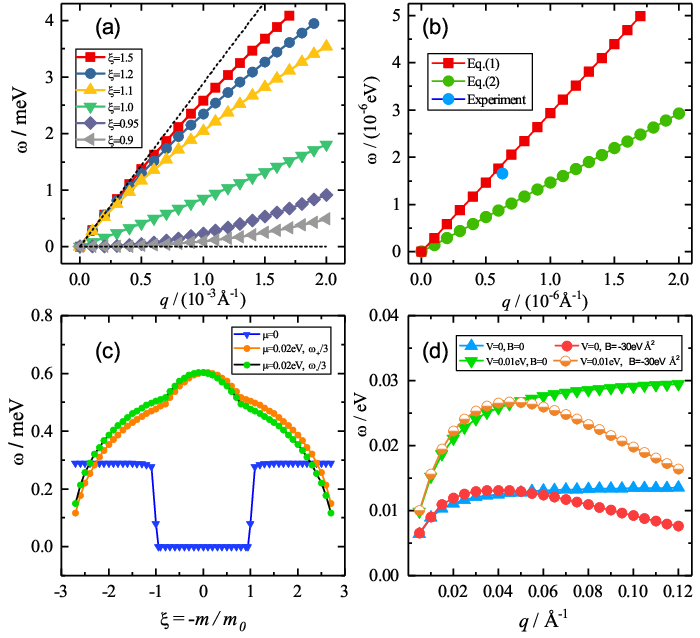}
\par\end{centering}
\centering{}\caption{(a) Energy dispersion of the unidirectional edge plasmon for selected
values of $\xi$. (b) The theoretical dispersions of the chiral edge
plasmon in the long-wavelength limit is in accordance with the experimental
results in Ref. \citep{Mahoney2017NC}. (c) The dependence of frequency
of the chiral edge plasmon with $q=10^{-4}/\mathbf{\mathrm{\mathring{A}}}$
on $\xi$ for different chemical potentials. (d) The velocity of the
edge plasmon in topologically nontrivial phase with $\xi=1.5$ becomes
negative for a sufficiently large wave vector $q$. The parameters
are $D=0$, and $V=0.01\mathrm{eV}$. \label{fig3} }
\end{figure}

We solve the matrix equation in Eq. (\ref{DetEq}) in the undoped
case and plot the dispersion curves of the edge plasmons in Figs.
\ref{fig3}(a)-\ref{fig3}(d) \citep{SoftBC}. When $m<0$, both Eq.
(\ref{DetEq}) and Eq. (\ref{AppEq}) have only one solution, in which
the wave vector $q$ is positive. So the edge plasmon is an unidirectional
mode and propagates in the positive direction, $i.e.$ $\omega_{+}$.
If $m>0$, the sign of Berry curvature changes and the unidirectional
edge plasmon will accordingly become mode $\omega_{-}$. This salient
character agrees with the recent experiment about magnetoplasmons
in a QAH insulator \citep{Mahoney2017NC}. In their experiment, the
frequency of the edge plasmon is about GHz, which is marked in the
insert of Fig. \ref{fig3}(b). Our numerical results show that the
velocity of the edge plasmon is about $2.93\:\mathrm{eV}\cdot\mathbf{\mathrm{\mathring{A}}}$
and well consistent with the experimental data $2.63\:\mathrm{eV}\cdot\mathbf{\mathrm{\mathring{A}}}$
\citep{Mahoney2017NC}. Figure \ref{fig3}(c) shows the dependence
of energy dispersions of the chiral plasmons on the exchange field
$m$ for $\mu=0\:\mathrm{eV}$ and $\mu=0.02\:\mathrm{eV}$. For the
Fermi level in the band gap, the frequency of chiral plasmons is independent
on the magnitude of exchange fields, which is well consistent with
the experimental result \citep{Mahoney2017NC}. On the contrary, the
frequency of chiral plasmons strongly depends on the magnitude of
magnetization for the lightly doped case. 

In the long-wavelength limit $q\rightarrow0$, we have $\sigma_{xx}\approx\sigma_{yy}\approx0$,
and $\sigma_{xy}\approx\pm e^{2}/h$ for the nontrivial phase ($h$
is Planck's constant and $e$ the electron charge). Inserting $\sigma_{xy}$
into Eq. (\ref{AppEq}), one immediately gets the dispersion relation of the edge plasmon: 
\begin{equation}
\omega_{\mathrm{H}+}\left(q\right)=\frac{q\sigma_{xy}}{2\sqrt{2}\epsilon}=\frac{|q|}{2\sqrt{2}\epsilon}\frac{e^{2}}{h}.\label{dispEP}
\end{equation}
It is clear that such edge plasmon is an acoustic mode whose velocity
is solely determined by the quantum Hall conductivity $e^{2}/h$ and
the effective dielectric constant of the environment, but independent
of the detailed parameters of the system. The corresponding velocity
of the edge plasmon in the long-wavelength limit is about $1.46\:\mathrm{eV}\cdot\mathbf{\mathrm{\mathring{A}}}$
and much less than the experimental value \citep{VelocityEMP}. This
is a unique collective mode appearing only in the QAH insulators.
The numerical solutions share the same character, as shown in Fig.
\ref{fig3}(a). When the system is in a nontrivial phase ($\xi>1$),
all the dispersion curves of the edge plasmons approach the same asymptotic
line in the long-wavelength limit. While for a trivial phase ($\xi<1$),
the dispersion curves tend to zero. Thus, the dispersion properties
of the unidirectional edge plasmon strongly depends on the topology
of the system. In sum, our work could reproduce all key features of
the chiral edge plasmons in the QAH insulator. It is one of the central
results in this work. 

\textit{Origin of negative dispersion.-{}-}Figure \ref{fig3}(d) shows
that, when the wave vector $q$ is large enough, the dispersion starts
to bend downward such that the group velocity of the edge plasmons
could become negative. It happens in both the topologically trivial
and nontrivial phases. In the rest of this section, we would like
to reveal the mechanism for the negative dispersion of edge plasmons. 

For simplicity, we assume $V=0$ and $D=0$, the effective Hamiltonian
in Eq. $\left(\ref{Ham}\right)$ reduces to two diagonalized blocks
with replacing $\Delta/2$ with $M_{\pm}=\left(m\pm\Delta\right)/2$
in $h_{\pm}(\boldsymbol{k})$. When $B=0$, the Hamiltonian above
describes 2D massive Dirac fermions with mass of $M_{\pm}$. By use
of the Kubo formula, one can calculate the transverse conductivity in the intrinsic case ($\omega>0$) \citep{SMCEP}
{\small{}
\begin{equation}
\sigma_{xy}(q,\omega)=\frac{e^{2}}{h}\sum_{a=\pm}\frac{M_{a}}{\sqrt{\upsilon_{F}^{2}q^{2}-\omega^{2}}}\left(\arctan\frac{2\left|M_{a}\right|}{\sqrt{\upsilon_{F}^{2}q^{2}-\omega^{2}}}-\frac{\pi}{2}\right),\label{sigxy}
\end{equation}
}where the exchange field $m$ gives rise to a non-zero transverse
conductivity. In the limit $q\rightarrow\infty$, we have $\sigma_{xy}=-e^{2}m\pi/2h\upsilon_{F}|q|$.
Consequently, from Eq. (\ref{AppEq}) one get a single unidirectional
edge plasmon, whose energy is independent of $q$: $\omega_{\mathrm{H}+}=\frac{e^{2}|m|\pi}{4\sqrt{2}h\epsilon\upsilon_{F}}$.
For $V\neq0$, this is also true suggested by the numerical results
(seen in Fig. \ref{fig3}(d)). Thus, for a large enough wave vector
$q$, the dispersion curve of the unidirectional edge plasmon (green
and blue lines) becomes a flat band, and no negative dispersion appears
in the system. 

When $\upsilon_{F}=0$, Eq. (\ref{Hpm}) reduces to a model for the
conventional 2D semiconductors. Accordingly, the integrand in the
Kubo formula becomes an odd function of $k_{x}$, and then one gets
$\sigma_{xy}(q,\omega)=0$. As a result, $\omega_{+}$ vanishes. More
generally, this is true even in the situation that $V$$\neq$0 and
$D$$\neq$0 so long as there is no term like $k_{x}^{n}$ in the
Hamiltonian where $n$ is an odd number. Thus, there is even no edge
plasmons in the conventional 2D semiconductors with a vanishing $\sigma_{xy}$. 

One thus finds that the negative dispersion of edge plasmons requires
the coexistence of the linear and quadratic terms of $k$. It is entirely
different from the previous mechanisms \citep{Aryasetiawan1994PRL,giuliani2005qtel,vanWezel2011PRL}.
In this situation, $\omega\propto q\sigma_{xy}$ is still satisfied
but the quadratic term $Bk^{2}$ will change the behavior of the transverse
conductivity. In the limit $q\rightarrow\infty$, our numerical calculation
implies that $\sigma_{xy}\varpropto|q|^{-\alpha}$ with $\alpha\approx4$.
As a result, the group velocity of the edge plasmon (red and orange
lines) becomes negative, as shown in Fig. \ref{fig3}(d). 

\textit{Manipulation of chiral plasmons}.\textit{-{}-}It has been
demonstrated that the transverse conductivity $\sigma_{xy}(q,\omega)$
plays a crucial role in forming the chiral plasmons \citep{Lin2020PRL}.
Thus, one could manipulate the chiral edge plasmons including the
number and the direction of the modes by tuning $\sigma_{xy}(q,\omega)$
in experiments. Specifically, the tunable Fermi level through the
gate voltage, ionic liquid or doping \citep{Zheng2021NC} could change
$\sigma_{xy}(q,\omega)$ even its sign. For instance, for the $n$-doped
case, the bigger the Fermi energy $\mu$, the smaller $\sigma_{xy}(q,\omega)$.
As a result, the energy split of the chiral plasmons $\Delta\omega\equiv\omega_{+}-\omega_{-}$
will decrease. On the other hand, higher Fermi energy can also increase
the gap between $\omega_{bulk}$ and $\omega_{+}$, preventing $\omega_{+}$
from merging with the bulk mode. 
\begin{figure}
\begin{centering}
\includegraphics[scale=0.6]{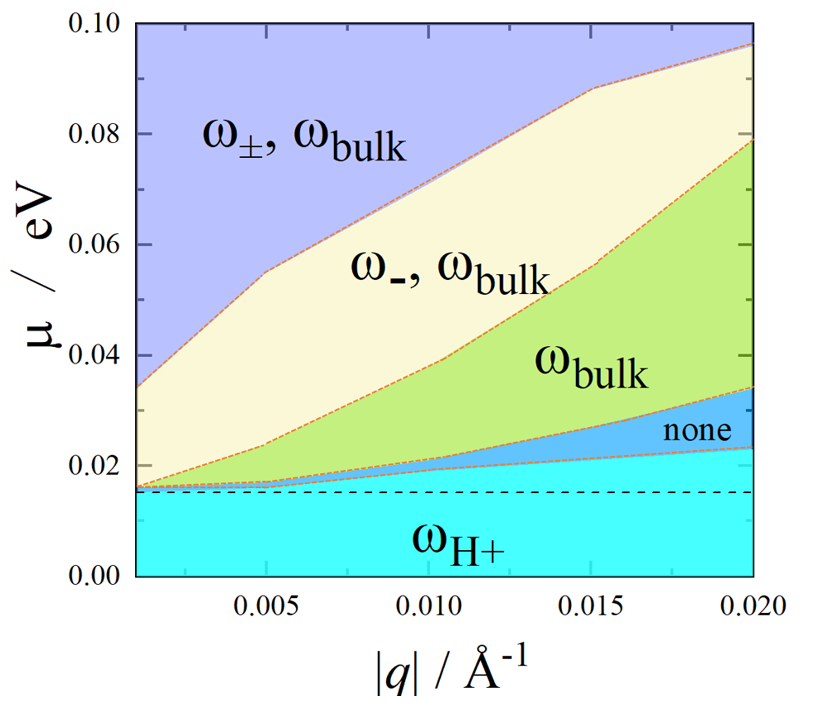}
\par\end{centering}
\centering{}\caption{Dependence of the bulk and edge plasmons on the Fermi energy and the
wave vector. The horizontal dashed line indicates the magnitude of
band gap. Parameters are identical to those in Fig. \ref{fig2}(a).
\label{fig4}}
\end{figure}

Figure \ref{fig4} depicts the behaviors of the bulk and chiral edge
plasmons versus the wave vector $q$ and the Fermi energy $\mu$.
For a larger enough Fermi energy $\mu$, the edge plasmons $\omega_{\pm}$
and the bulk plasmon $\omega_{bulk}$ coexist. As $\mu$ decreases,
$\omega_{+}$ mode becomes delocalized along the direction perpendicular
to the edge and gradually merges with the bulk plasmon. Then the $\omega_{-}$
mode and the bulk mode $\omega_{bulk}$ will enter the intra-band
SPE region successively and are damped, leaving no plasmon excitations.
Note that this always happens for a general wave vector. It is worth
pointing out that when the Fermi level lies in the band gap, there
exists only one single edge-plasmon mode $\omega_{\mathrm{H}+}$,
which is dominated by the transverse conductivity. Meanwhile, when
the system is lightly doped, $\omega_{\mathrm{H}+}$ can also survive
until the longitudinal conductivity becomes comparable to the transverse
one. 

In summary, we showed that the Berry curvature splits the degeneracy
of edge plasmons, leading to the chiral plasmons in the QAH insulators.
When the system enters into the QAH phase, only one unidirectional
edge plasmon survives, whose propagation direction can be flipped
by the external magnetization. In the long-wavelength limit, the unidirectional
edge plasmon is acoustic and entirely determined by the anomalous
Hall conductivity and the effective dielectric constant. We provide
a quantitative explanation of the recent observation of the chiral
edge plasmons in QAH insulators and find the negative dispersion of
chiral edge plasmon and reveal its unusual physical origin. Therefore,
our work could be applicable to the tunable intrinsic magnetic topological
materials hosting QAHE, such as $\mathrm{Mn}\mathrm{Bi}_{2}\mathrm{Te}_{4}$
\citep{Gong2019CPL,Deng2020science,Mei2024PRL,Chong2024PRL}. 

\textcolor{black}{This work was supported by the National Key R\&D
Program of China (Grant No. 2020YFA0308800)} and by the National Natural
Science Foundation of China (No.\textcolor{black}{{} 12174394}). J.Z.
was supported by the HFIPS Director\textquoteright s Fund (Grant Nos.
YZJJQY202304 and BJPY2023B05) and Chinese Academy of Sciences under
contract No. JZHKYPT-2021-08 and also by \textcolor{black}{the High
Magnetic Field Laboratory of Anhui Province.}

\bibliographystyle{apsrev4-1}
\bibliography{CEPlasmon}

\begin{thebibliography}{66}
\bibitem [{Vel()}]{VelocityEMP}%
  \BibitemOpen
\bibfield  {journal} {  }\href@noop {} {\bibinfo  {journal} {The discrenpency
  between theoretical values of the velocity of the chiral edge plasmon through
  the numerical and approximate methods and the experimental result might
  originate from the treatment of the density profile of electrons near the
  edge of the QAH insulator}\ }\BibitemShut {NoStop}%
\end{thebibliography}

\end{document}